\documentclass[fleqn,usenatbib]{mnras}

\usepackage{newtxtext,newtxmath}

\usepackage[T1]{fontenc}

\DeclareRobustCommand{\VAN}[3]{#2}
\let\VANthebibliography\thebibliography
\def\thebibliography{\DeclareRobustCommand{\VAN}[3]{##3}\VANthebibliography}

\usepackage{graphicx}	
\usepackage{amsmath}	
\usepackage{xcolor} 
\usepackage{multirow} 

\usepackage{booktabs}   
\usepackage{multirow}   
\usepackage{array}

\title[Age bimodality and bar-driven evolution ]{Age bimodality in pseudo-bulges of barred spiral galaxies: Bar-driven evolution across cosmic time}

\author[Kavin Kumar et al.]{
Kavin Kumar N. R.,$^{1}$\thanks{E-mail: kavin.nagarajanrajkumar@research.uwa.edu.au (KK)}
Saili Keshri,$^{2,3}$\thanks{E-mail: saili.keshri@iiap.res.in (SK)} and Sudhanshu Barway$^{2}$\thanks{E-mail: sudhanshu.barway@iiap.res.in (SB)}
\\
$^{1}$ICRAR, The University of Western Australia, 35 Stirling Highway, Crawley, WA 6009, Australia\\
$^{2}$Indian Institute of Astrophysics (IIA), II Block, Koramangala, Bengaluru 560 034, India\\
$^{3}$Department of Physics, Pondicherry University, R.V.Nagar, Kalapet, Puducherry, India, 605014
}

\date{Accepted XXX. Received YYY; in original form ZZZ}

\pubyear{\the\year{}}

\begin{document}
\label{firstpage}
\pagerange{\pageref{firstpage}--\pageref{lastpage}}
\maketitle

\begin{abstract}
We investigate the stellar population properties of pseudo-bulges in barred galaxies drawn from the Sloan Digital Sky Survey (SDSS DR7) to assess how bars regulate central star formation and secular evolution. Our sample comprises barred spiral and barred lenticular (S0) galaxies with reliable spectroscopic indices obtained from multicomponent structural decompositions. Stellar ages and recent star formation are traced using the 4000\AA\ break strength ($D_{n}(4000)$) and the Balmer absorption index ($H\delta_{A}$), complemented by bulge, bar, and disc colours. Barred spirals show a clear bimodality in $D_{n}(4000)$, with peaks at $D_{n}(4000)\sim1.3$ and $\sim1.8$. Low-$D_{n}(4000)$ pseudo-bulges exhibit strong $H\delta_{A}$ absorption, blue colours, and high specific star-formation rates, indicating young, actively growing centres. High-$D_{n}(4000)$ systems instead show weak $H\delta_{A}$, red colours, and low sSFR, consistent with older, quenched pseudo-bulges. Barred S0s display an old-bulge-dominated distribution, suggesting that gas-poor barred spirals transition into S0s following disc-wide quenching. We also find elevated AGN incidence among old pseudo-bulges. These trends support a scenario in which bars funnel gas inward to build pseudo-bulges and later suppress central star formation by depleting or stabilising the inflow. IFU observations show that bars assemble cold nuclear discs that age and quench over time, while high-redshift imaging confirms that bars are already present at $z\sim4$, implying that this evolutionary cycle operates across cosmic time. The strong correspondence between stellar age, colour, and structure indicates that bar-driven secular evolution governs both the growth and quenching of central components, linking blue barred spirals to red S0 galaxies.

\end{abstract}




\section{Introduction}
\label{sec:intro}
The formation and evolution of galactic bulges remains a central problem in understanding the growth of disc galaxies, as bulges display a wide range of structural and stellar population properties that reflect multiple formation pathways. Bulges are typically separated into two main classes: \emph{classical bulges}, which are centrally concentrated, dynamically hot, and dominated by pressure support, and \emph{disc-like (pseudo-)bulges}, which are more rotationally supported, exhibit flattened structures and ongoing star formation, and often retain disc-like morphologies. This distinction is critical, as it implies fundamentally different evolutionary histories. Classical bulges are generally thought to form rapidly through major mergers \citep{hammer05, Hopkins10b} or clump-collapse through violent internal instabilities \citep{Elmegreen08, Ceverino15}. At the same time, satellite accretion during minor mergers can also contribute to their growth from disc material \citep{Aguerri01,Eliche-Moral06}.
In contrast, pseudo-bulges arise gradually through long-term, internal (secular) processes within discs \citep{kormendy04,gadotti09}. Their formation is closely tied to the dynamical evolution of bars, which are ubiquitous in disc galaxies and efficiently redistribute angular momentum, driving gas inward to fuel central star formation \citep{Aguerri09}. However, pseudo-bulges may also form through external processes, such as the accretion of low-mass satellites \citep{Eliche-Moral11}, indicating that their evolutionary origin may be diverse as well.

In the nearby Universe, around 70\,per cent of disc galaxies host stellar bars \citep{deVaucouleurs63, Knapen00, nair10, Masters11, Salo2015}. Bars are dynamically stable structures that drive gas from the outer disc toward the central kiloparsec through bar-induced gravitational torques \citep{Combes81, Debattista04, kormendy04, Athanassoula13}. This inflow enhances central star formation and can also feed active galactic nuclei (AGN) \citep{Athanassoula1992, Ho97, Coelho11, Ellison11}. Bar-driven gas transport is therefore a key mechanism for building pseudo-bulges through sustained, central star formation \citep{kormendy04, Athanassoula03, Fisher16, hopkins10}, a result supported by hydrodynamical simulations that show bars increase central mass concentration and alter bulge dynamics \citep{Fanali2015, Spinoso17, Neumann2020}.

However, gravitational torques and resonances may also stabilise or deplete inflowing gas, ultimately suppressing further star formation \citep{frosst25}. This dual role, fueling growth and later inducing quenching, positions bars as powerful agents of secular transformation in disc galaxies. Observational support for this two-phase behaviour is growing: bars are more common in redder and more massive hosts \citep{Masters11,vera16}, and bar strength correlates with older stellar populations and lower specific star-formation rates \citep{Spinoso17, James18}, suggesting that bars may accelerate the transition from the star-forming main sequence to the green valley. Understanding this apparent contradiction requires examining how bar-driven processes influence the ages of bulge stars across diverse morphologies and evolutionary stages.

AGN feedback may further regulate central growth. Negative feedback from AGN activity can deplete gas within the inner kiloparsec and suppress bar-driven star formation \citep{Ellison11, Lammers23}, and recent work \citep{Lammers23, Keshri25} shows that such suppression is confined to the central region, without strongly affecting galaxy-wide star formation.

Pseudo-bulges provide an excellent laboratory for disentangling these processes, as they retain the dynamical imprint of their bar-driven formation. Their stellar populations encode the timing and efficiency of gas inflow, central star formation, and subsequent quenching, allowing the evolutionary role of bars to be assessed directly. Previous work has reported a clear bimodality in the stellar ages of pseudo-bulges in unbarred S0 galaxies \citep{mishra17b}, interpreted as evidence for multiple evolutionary channels. However, whether a similar or fundamentally different behaviour is present in barred systems where bar-driven inflow and circumnuclear star formation remain active has not yet been established. In particular, the role of bars in shaping the growth and quenching of pseudo-bulges and in driving the evolutionary connection between barred spirals and barred S0 galaxies remains poorly constrained.

It is important to distinguish the scope of this work from previous studies of barred galaxies. For example, \citet{Barway20} examined the photometric properties of bars in disc galaxies and found that bluer bars are more common in a subset of S0 galaxies than in spirals, suggesting possible rejuvenation of star formation in S0 bars. However, the study did not address the stellar population properties of bulges (classical or pseudo) or their connection to bar-driven secular evolution. Similarly, while previous work (e.g. \citet{mishra17a}) has focused on unbarred S0 galaxies, the stellar population properties of pseudo-bulges in \emph{barred spiral and S0 galaxies} have not been systematically explored.

In this paper, we investigate the stellar populations of pseudo-bulges in barred galaxies using a large, homogeneous sample drawn from the Sloan Digital Sky Survey Data Release~7 (SDSS~DR7), combined with structural decompositions from \citet{Kruk18}. We use spectral indices such as the 4000\,\AA\ break strength ($D_{n}(4000)$) and the Balmer absorption index ($H\delta_{A}$) to trace mean stellar ages and recent star-formation histories, complemented by bulge, bar, and disc colours. This approach allows us to assess the diversity of pseudo-bulge stellar populations in barred galaxies and to examine the role of bar-driven secular processes, to quantify how this links to ongoing or quenched central star formation, and evaluate whether bar-driven evolutionary processes can explain the observed connection between spirals and S0 galaxies.

The structure of this paper is as follows. In Section~\ref{sec:data}, we describe the data sources, sample selection, and morphological classification scheme used to obtain the initial sample. In Section~\ref{sec:pb}, we explain the methods used to further narrow down on the pseudo-bulge hosting galaxies out of this sample. In Section~\ref{sec:results}, we present our analysis of stellar age indicators, colours, and specific star-formation. In Section~\ref{sec:summary} we discuss the implications of our findings for bar-driven secular evolution and quenching along with the concluding remarks. Throughout the paper, we use the WMAP Seven-Year Cosmological parameters \citep{wmap11} with $(\Omega_{M}, \Omega_{\Lambda}, h)$ = (0.27, 0.73, 0.71). 

\begin{figure}
\includegraphics[scale = 0.45]{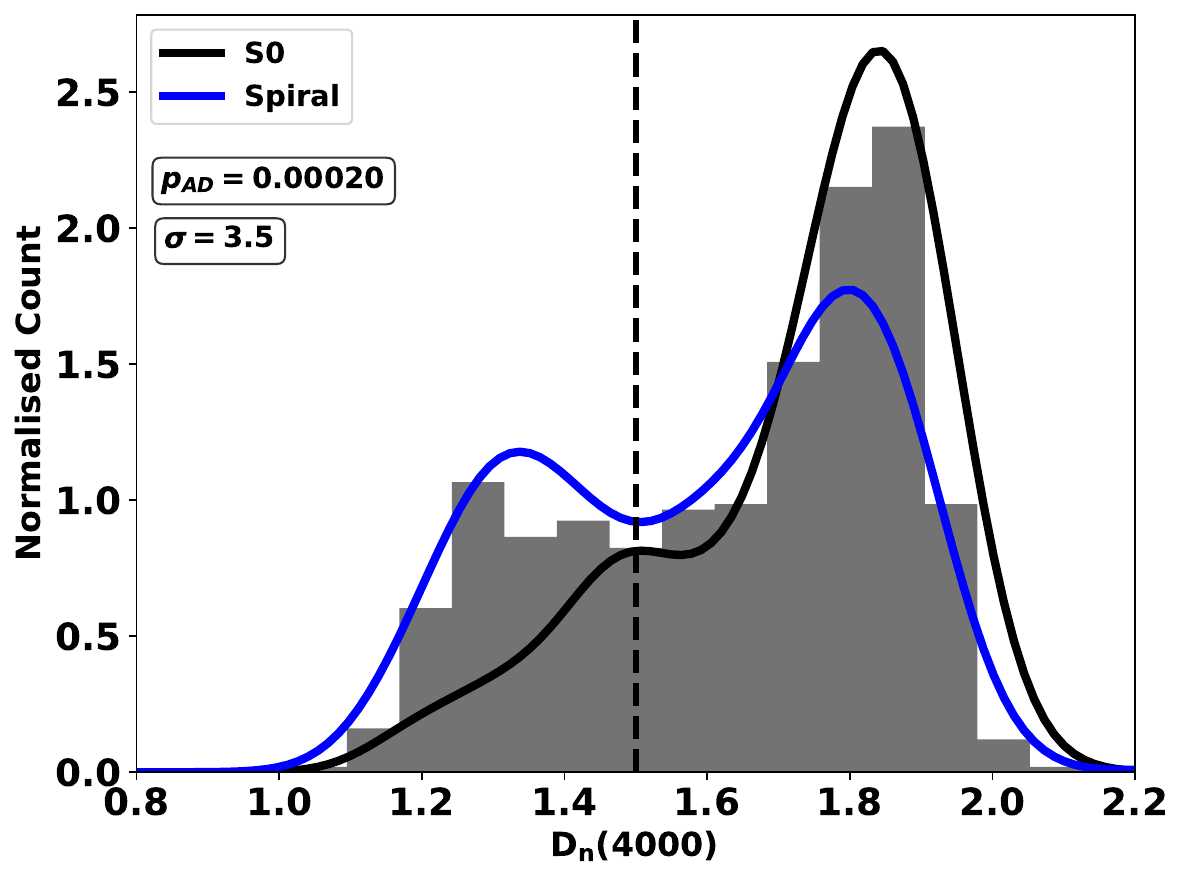}
\caption{Distribution of $D_{n}(4000)$ index for 677 barred galaxies. Blue and black curves show the Gaussian kernel–smoothed distributions for barred spirals and barred S0s, respectively. The black vertical dashed line marks the conventional division between younger, star-forming systems and older, quenched stellar populations.}
\label{fig:dn4000}
\end{figure}

\section{Data and Sample selection} 
\label{sec:data}
Our analysis is based on the catalogue of barred galaxies presented by \citet{Kruk18}, which provides a comprehensive two-dimensional (2D) multi-band photometric decomposition for $\sim$3500 nearby barred systems drawn from the Sloan Digital Sky Survey (SDSS; DR7). The decomposition was performed using the \textsc{galfitm} code \citep{Haubler2013, vika2013}, a wavelength-dependent extension of \textsc{galfit} \citep{Peng2010} developed within the \textsc{MegaMorph} project. Unlike traditional single-band fitting, this method leverages the full wavelength coverage of the SDSS ugriz filters to improve the signal-to-noise ratio and parameter accuracy. Each galaxy was modelled in the five SDSS bands ($ u$, $ g$, $ r$, $ i$, $ z$) simultaneously, allowing for consistent structural parameters and improved signal-to-noise across filters. In these models, the Sérsic index ($n$) and effective radius ($r_e$) for each component (disc, bar, and bulge) were constrained to be constant across all five bands, while magnitudes were allowed to vary freely to determine component colours. To ensure robust convergence, the fits followed an iterative process: starting with a single-Sérsic fit, followed by a disc-bar model, and concluding with a disc-bar-bulge decomposition where visually appropriate. The fitting adopted a S\'ersic profile for the bulge and bar components and an exponential profile for the disc, yielding magnitudes, effective radii, and S\'ersic indices for each component. All final models were subjected to rigorous visual inspection to confirm that the components corresponded to genuine physical structures. The resulting parameters, corrected for Galactic extinction and $k$-corrections, are reliable for the $g$, $r$, and $i$ bands, which are used throughout this work.

The uncertainties quoted by \textsc{galfitm} are primarily statistical and do not include systematic effects from sky background, point–spread function (PSF) modeling, or parameter degeneracies. To minimise dust–related errors, we restrict the analysis to galaxies with inclination $i \le 60^{\circ}$, ensuring that the majority of the sample is close to face-on. Typical median S\'ersic indices and component colours from \citet{Kruk18} are consistent with a bar-dominated population of late-type galaxies.

\begin{figure}
\centering
\includegraphics[scale = 0.45]{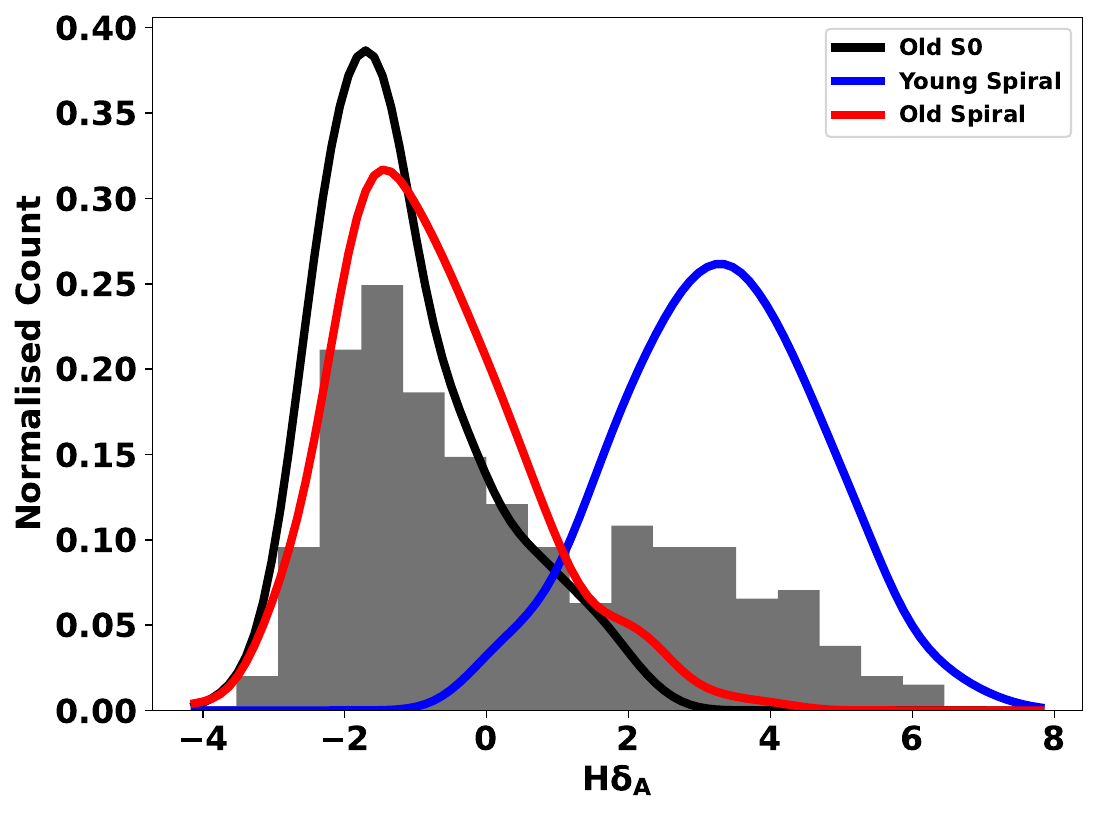}
\caption{Distribution of $H\delta_{A}$ for 677 barred galaxies. The Gaussian kernel–smoothed distributions for OB–Sp, YB–Sp, and barred S0 hosts are shown in red, blue, and black, respectively.}
\label{fig:hda}
\end{figure}

\subsection{Aperture selection and fibre coverage}
Spectroscopic indices such as $D_{n}(4000)$ and $H\delta_{A}$ are taken from the MPA–JHU SDSS database \citep{kauffmann03}, measured within the 3~arcsec diameter fibre aperture. Because the physical region covered by the fibre depends on redshift, contamination from bar and/or disc light could bias the bulge measurements. To minimise this effect, we include only galaxies for which the fibre diameter is smaller than twice the bulge effective radius ($r_{\mathrm{fibre}} < 2\,r_{e,{\mathrm b}}$). This criterion ensures that the bulge dominates the observed spectrum. Out of the $\sim$2200 barred galaxies with three-component fits in \citet{Kruk18}, 1600 galaxies satisfy this condition and form the parent sample for our analysis. 

\begin{figure*}
\includegraphics[scale = 0.35]{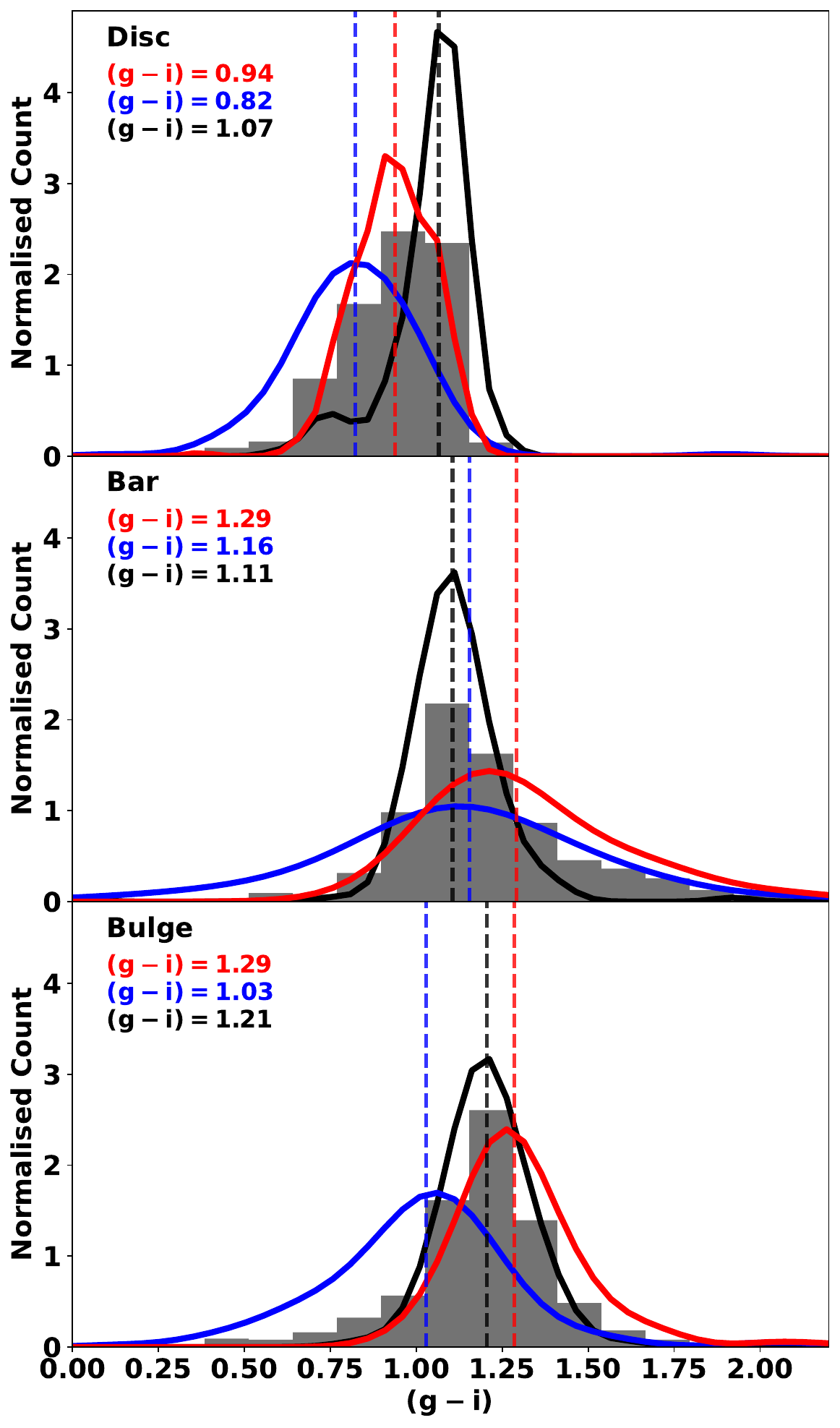}
\includegraphics[scale = 0.35]{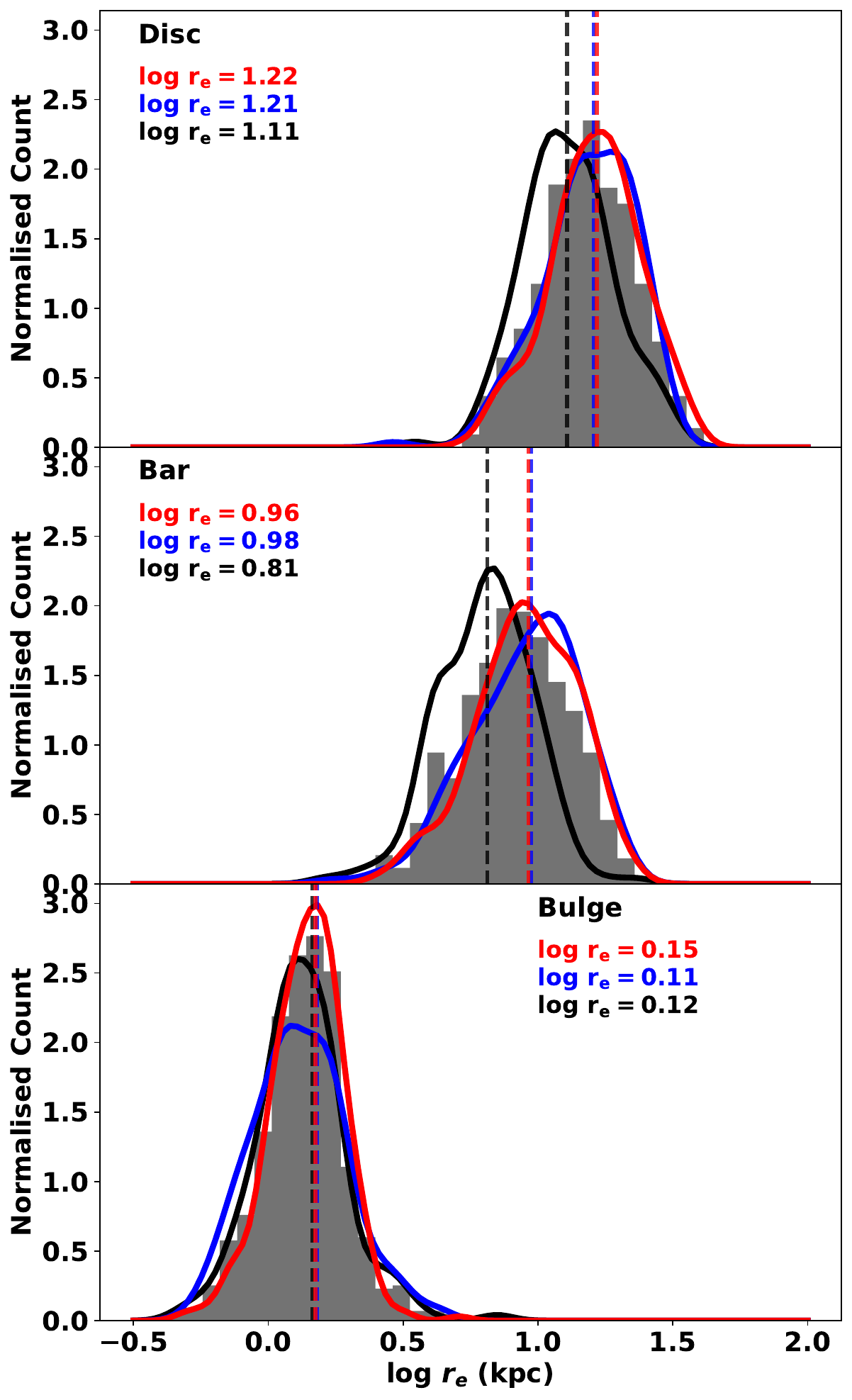}
\caption{\textbf{Left:} $(g-i)$ colour distributions for the bulge, bar, and disc components of 677 barred galaxies. 
\textbf{Right:} Distributions of $\log R_{\mathrm{e}}$ for the corresponding bulge, bar, and disc effective radii. In both panels, the Gaussian kernel–smoothed distributions for OB–Sp, YB–Sp, and OB–S0 hosts are shown in red, blue, and black, respectively. Vertical dashed lines mark the median values for each component and parameter.}
\label{fig:col_size}
\end{figure*}


\subsection{Morphological classification}
To separate spiral and lenticular systems, we cross-matched the barred galaxy catalogue with the morphological probabilities from \citet{Huertas-Company11}, who provide $P(E)$, $P(S0)$, $P(Sab)$, and $P(Scd)$ for SDSS galaxies based on a support-vector machine classifier. Following \citet{Meert15}, we converted these probabilities into a continuous T–type using their linear relation (their equation~7) and classified galaxies with $T \le 0.5$ as S0s and those with $T > 0.5$ as spirals. The cross-match was performed using SDSS object identifiers with a positional tolerance of $3''$.

After applying the inclination, fibre coverage, and morphological criteria, our final working sample comprises 697 barred S0 galaxies and 859 barred spirals. For each galaxy, we use the photometric parameters (magnitudes, colours, S\'ersic indices, and effective radii) from \citet{Kruk18}, and the spectroscopic measurements ($D_{n}(4000)$, $H\delta_{A}$ and emission line fluxes) are taken from the MPA–JHU catalogue. We have taken the stellar mass ($M_{*}$), and star-formation rates (SFRs) from GALEX-SDSS-WISE LEGACY CATALOG (GSWLC - X2) \citep{salim18}. The local environmental density, used in Section ~\ref{sec:sf_activity} is taken from \citet{baldry06}, who compute the surface galaxy density $\Sigma_{N}=N/(\pi d_{N}^{2})$ based on the projected distance to the fifth nearest neighbour within a velocity window of $\pm1000~{\rm km\,s^{-1}}$.
This carefully curated dataset provides a statistically robust sample of barred disc galaxies with well-constrained structural and stellar-population parameters, enabling us to examine how bar-driven secular evolution influences the ages of pseudo-bulges in different morphological types.


\section{Identifying pseudo-bulges}
\label{sec:pb}
Bulges in disc galaxies are broadly classified into two categories: \textit{classical bulges}, which are dynamically hot, spheroidal systems formed through rapid processes such as mergers or monolithic collapse \citep{Elmegreen08}, and \textit{pseudo-bulges}, which are rotationally supported components that arise gradually through internal secular evolution \citep{kormendy04, Fisher16}. Because stellar bars are known to drive gas inflows and redistribute angular momentum, the presence of a bar is often linked to the growth of pseudo-bulges. Distinguishing between these two bulge types is therefore essential for evaluating the impact of bar-driven secular evolution on the central stellar populations of galaxies.

Several structural and kinematic diagnostics have been proposed in the literature to identify pseudo-bulges \citep[e.g.][]{gadotti09, fisherdrory10, Weinzirl2009, MendezAbreu2017}. A single criterion, such as the S\'ersic index, is insufficient on its own, since observational uncertainties and projection effects can blur the distinction between bulge types. To obtain a robust classification, we adopt a conservative, multi-parameter approach that combines three independent criteria—morphological, dynamical, and photometric—similar to the method of \citet{mishra17a}. A galaxy is considered to host a pseudo-bulge if it simultaneously satisfies the following conditions:

\begin{enumerate}
    \item[(i)] The bulge lies below the $3\sigma$ lower boundary of the Kormendy relation (KR) for elliptical galaxies, defined as
    \begin{equation}
        \langle\mu_{b}(<r_{e})\rangle = (2.330 \pm 0.047)\,\log(r_{e}) + (18.160 \pm 0.024),
    \end{equation}
    where $\langle\mu_{b}(<r_{e})\rangle$ is the mean surface brightness within the bulge effective radius $r_{e}$, measured in the SDSS $r$ band. The rms scatter about this relation is $0.429$\,mag\,arcsec$^{-2}$ \citep{mishra17a}. Bulges lying more than $3\sigma$ below the best-fitting line are classified as pseudo-bulges. This criterion effectively separates systems with low surface brightness and extended profiles from the high-concentration classical bulges that follow the elliptical sequence.

    \item[(ii)] The central stellar velocity dispersion, $\sigma_{0}$, is less than $130~{\rm km~s^{-1}}$. This threshold, supported by \citet{Fisher16}, distinguishes dynamically cold, rotation-dominated pseudo-bulges from pressure-supported classical bulges that typically exhibit higher velocity dispersions.

    \item[(iii)] The bulge S\'ersic index satisfies $n_{\mathrm{b}} < 2$. Numerous studies have shown that exponential or near-exponential profiles well represent pseudo-bulges, while classical bulges exhibit $n_{\mathrm{b}} \gtrsim 3$ \citep{kormendy04, gadotti09}.
\end{enumerate}

Only galaxies satisfying all three conditions are retained as secure pseudo-bulge hosts. Applying these criteria to our barred sample yields a total of 677 pseudo-bulge galaxies, comprising 203 barred S0s and 474 barred spirals. This fraction ($\sim$42\%) is somewhat lower than the $\sim$60–75\% pseudo-bulge fraction commonly reported for barred galaxies \citep[e.g.][]{Laurikainen2004, Salo2015, MendezAbreu2017}, reflecting our intentionally conservative selection strategy. These galaxies lie in the redshift range 0.006–0.06, corresponding to luminosity distances of approximately 25–265 Mpc. The redshift range of our sample implies that the fixed angular size of the SDSS fibre probes different physical scales, potentially introducing aperture-related systematics, particularly for nearby galaxies where the fibre primarily samples the central regions. To quantify this effect, we compute the ratio of fibre radius to $r_{e,bulge}$ in the lowest ($\le$ 0.02) and highest redshift ($\ge$ 0.04 ) ranges. The median value of this ratio suggests that fibre covers $\sim$56\% of the $r_{e,bulge}$ in the low-redshift galaxies, whereas it covers $\sim$84\% of $r_{e,bulge}$ in the high-redshift galaxies, indicating that the fibre remains predominantly sensitive to the bulge region across the full redshift range.  Importantly, the low-redshift galaxies in our sample comprise only 13\% of the total sample. We therefore expect that aperture effects are unlikely to be the primary driver of the trends reported in this work.

It is important to note that \citet{Kruk18} report a mild underestimation of bulge S\'ersic indices ($\sim$30\%) in their multi-band fits due to image stacking and resolution effects. Consequently, some marginal classical bulges may scatter into the pseudo-bulge regime. However, since our classification additionally requires both the Kormendy and $\sigma_{0}$ criteria to be met, the impact of this bias is minimal. Visual inspection of a random subset confirms that the adopted thresholds effectively isolate low-concentration, disc-like bulges typically associated with secular evolution.

The use of multiple diagnostics not only reduces contamination but also ensures consistency with previous SDSS-based studies of barred and unbarred systems. For instance, \citet{mishra17b} employed similar criteria to analyse unbarred S0s and spirals, finding that pseudo-bulges in S0s exhibit a distinct age bimodality in $D_{n}(4000)$. Our study extends this framework to barred galaxies, enabling a direct comparison of stellar population trends and the role of bars in accelerating or suppressing central star formation.

In the following section, we explore how the stellar populations of these pseudo-bulges vary across different host morphologies. In particular, we examine whether barred spiral galaxies display a bimodal distribution in $D_{n}(4000)$, an indicator of mean stellar age, and how this behaviour contrasts with that of barred S0s.


\begin{figure*}
\centering
\includegraphics[scale = 0.27]{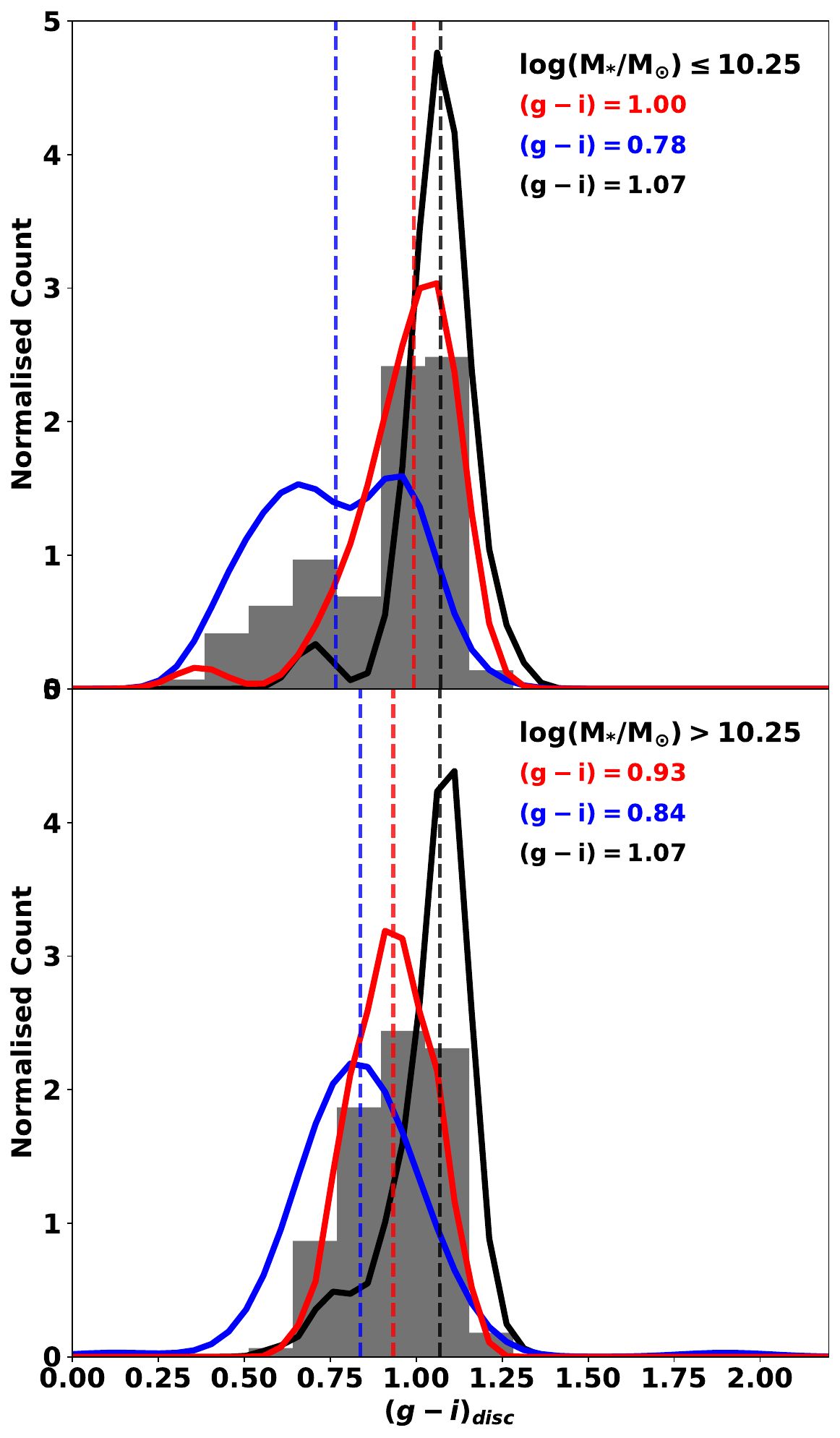}
\includegraphics[scale = 0.27]{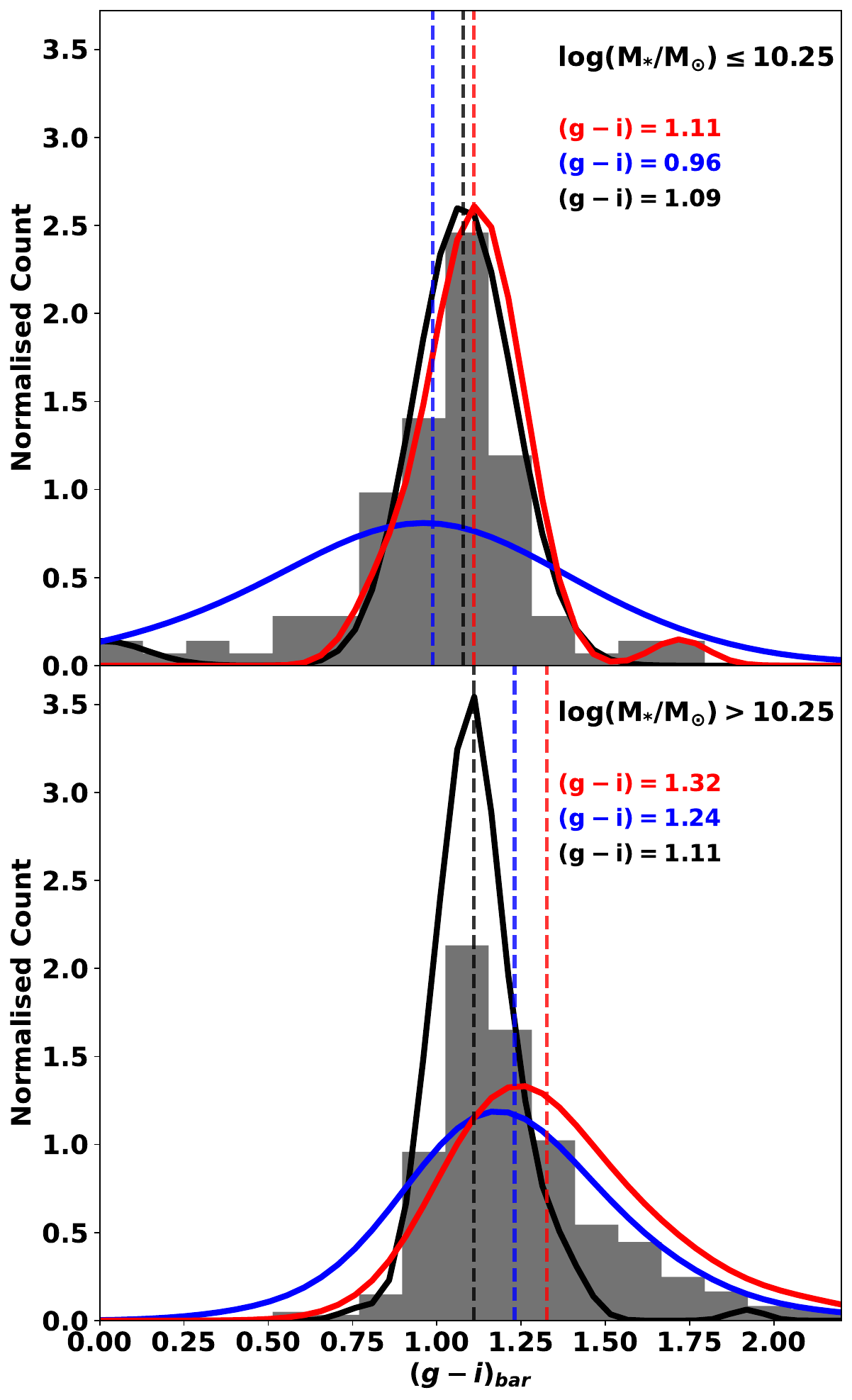}
\includegraphics[scale = 0.27]{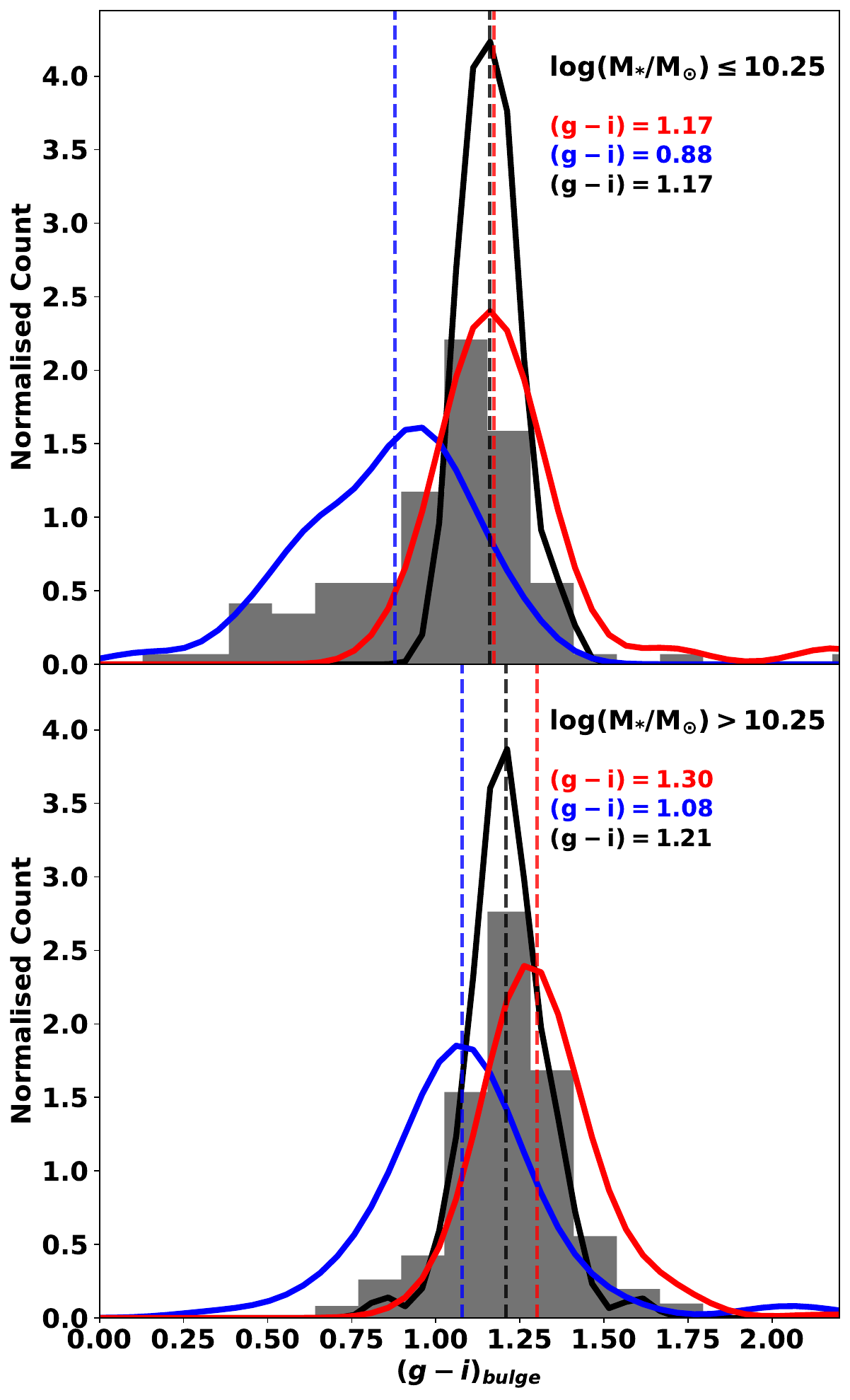}
\caption{$(g-i)$ colour distributions for the disc (left), bar (centre), and bulge (right) components, shown separately for two stellar-mass bins. The top row corresponds to galaxies with $M_{*} \leq 10^{10.25}\,M_{\odot}$, while the bottom row shows systems with $M_{*} > 10^{10.25}\,M_{\odot}$. Gaussian kernel–smoothed distributions for OB–Sp, YB–Sp, and OB–S0 hosts are plotted in red, blue, and black, respectively. Vertical dashed lines mark the median colour of each morphological class in each panel.}
\label{fig:mass_cut_col}
\end{figure*}

\section{Results}
\label{sec:results}
\subsection{Age bimodality in pseudo-bulges of barred spirals}
\label{sec:agebimodality}
Different bulge formation mechanisms imprint distinct stellar population signatures in the centres of disc galaxies. Classical bulges, assembled rapidly through mergers or monolithic collapse, generally host old and metal-rich stars. In contrast, pseudo-bulges, formed gradually through internal secular processes, often retain ongoing or recent star formation. To probe how these mechanisms manifest in barred systems, we analyse the stellar population ages of our sample using two complementary spectroscopic diagnostics: the 4000 $\AA$ break strength ($D_{n}(4000)$) and the Balmer absorption index ($H\delta_{A}$). Together, these indices trace the mean age of the stellar population and recent star-formation activity of the pseudo-bulges.

The $D_{n}(4000)$ index measures the flux ratio across the 4000 $\AA$  break, which arises from the accumulation of metal absorption lines in the atmospheres of cool, low-mass stars \citep{Balogh99}. Young stellar populations dominated by hot O-B stars exhibit weak breaks ($D_{n}(4000)\lesssim1.3$), while old, quiescent systems display strong breaks ($D_{n}(4000)\gtrsim1.8$). \citet{kauffmann03} reported that intermediate values around $D_{n}(4000)\simeq1.5$ correspond to light-weighted mean ages of $\sim$2 Gyr. We present the $D_{n}(4000)$ distributions for barred S0 (black) and spiral (blue) galaxies hosting pseudo-bulges in Figure~\ref{fig:dn4000}. The barred spirals exhibit a striking bimodality with peaks at $D_{n}(4000)\approx1.3$ and $\approx1.8$, whereas S0s show a distribution skewed towards the older populations centred above $D_{n}(4000)\sim1.6$. To quantify these facts, we perform a Gaussian mixture model (GMM) decomposition (for detailed mathematical procedure, see \citet{Ness13}) for both the morphologies. Model selection is based on the Bayesian Information Criterion (BIC), evaluated for one to five components. For the barred spiral, the BIC reaches a minimum for a two-component model (-129.72) with a $\Delta$ BIC $\approx$ 110 lower than the single-component model (-19.60), indicating decisive evidence for bimodality. The low $D_{n}(4000)$ component ($\sigma$ = 0.207) encompasses 58\% of the probability density of the model, while the high $D_{n}(4000)$ component ($\sigma$ = 0.082) encompasses the remaining 42\% confirming the bimodality for the bulges of spiral galaxies. On the other hand, for S0 galaxies, the BIC also favours a two-component model (-142.53) with $\Delta$ BIC $\approx$ 66 relative to a single-component model (-76.15). In this case, low ($\sigma$ = 0.146 dex) and high ($\sigma$ = 0.082 dex) $D_{n}(4000)$ components account for 30\% and 70\% of the relative weights, respectively.

The probability distribution is dominated by high $D_{n}(4000)$ values, while the lower $D_{n}(4000)$ component shows only a weak peak around $\approx$1.5 (which corresponds to an intermediate-age stellar population rather than a young population). This behaviour is best described as a strongly skewed distribution toward older stellar populations (i.e., higher $D_{n}(4000)$). 

To quantify this difference, we divide the barred spirals at $D_{n}(4000)=1.5$, classifying systems with lower and higher values as hosting \textit{young} pseudo-bulges (YB) and \textit{old}  pseudo-bulges (OB), respectively. Among 474 barred spirals, 171 ($36$\%) host YB and 303 ($64$\%) host OB. In comparison, $\sim$90\% of barred S0s possess only old pseudo-bulges. The $D_{n}$4000 distributions of S0 and spiral galaxies differ significantly according to the Anderson-Darling test with a $p_{AD} = 0.0002$, a statistically signiﬁcant result (3.54 $\sigma$).

The observed age bimodality in barred spirals provides strong evidence for bar-driven quenching of pseudo-bulges. We note that pseudo-bulges in unbarred spiral galaxies have been reported to exhibit largely unimodal age distributions, whereas S0 galaxies show clear evidence of bimodality \citep{mishra17b}. In this context, the behaviour observed for barred spirals offers an important point of comparison for understanding how bars and galaxy morphology influence the evolution of pseudo-bulge stellar populations, and the strong bimodality seen only in barred spirals proves that the presence of a bar introduces a new, efficient quenching channel that operates specifically on the pseudo-bulge. Bars are known to redistribute angular momentum and funnel gas into the central regions \citep{Athanassoula03}, triggering enhanced star formation that contributes to pseudo-bulge growth. As the available gas is subsequently depleted or dynamically stabilised, central star formation declines, leading to the formation of older stellar populations. The coexistence of young and old pseudo-bulges in barred spirals can therefore be interpreted as reflecting different evolutionary stages of bar-driven secular processes.

The $H\delta_{A}$ index offers an independent probe of this process. It measures the equivalent width of the Balmer H$\delta$ absorption line, which is strongest in A-type stars and thus sensitive to star formation within the past $\lesssim1$\, Gyr \citep{kauffmann03}. High $H\delta_{A}$ values indicate post-starburst or recently active populations, whereas smaller values of $H\delta_{A}$ suggest that the galaxy is less likely to have had significant bursts of star formation in the last Gyr. Figure~\ref{fig:hda} shows the $H\delta_{A}$ distributions for the same barred pseudo-bulge sample. The correspondence between Figures~\ref{fig:dn4000} and~\ref{fig:hda} is evident, i.e. galaxies with low $D_{n}(4000)$ display strong $H\delta_{A}$ absorption, whereas those with high $D_{n}(4000)$ exhibit weak lines. The two indices are therefore tightly anti-correlated, consistent with previous SDSS-based analyses of star-formation histories \citep[e.g.][]{kauffmann03}. Quantitatively, young pseudo-bulge host spirals (YB-Sps) ($D_{n}(4000)<1.5$) have a median $H\delta_{A}\approx3.34$ \AA, compared with $\approx -0.93$ \AA for old pseudo-bulge host spirals (OB-Sps), confirming that the former retain significant intermediate-age stellar components. In contrast, barred S0s show consistently low $H\delta_{A}$ ($\approx -1.42$ \AA), indicating negligible star formation in the past $\sim1$ Gyr.

These trends demonstrate that the bimodality in $D_{n}(4000)$ truly reflects an age separation, rather than a mere metallicity effect. The coupling of low $D_{n}(4000)$ with strong $H\delta_{A}$ absorption provides direct evidence for residual star formation in the pseudo-bulges of barred spirals, while the high-$D_{n}(4000)$ \& low-$H\delta_{A}$ population represents quenched pseudo-bulges. The combined diagnostic therefore delineates a clear evolutionary sequence: bar formation triggers an inflow-driven central starburst, producing a young pseudo-bulge. As the gas reservoir is exhausted and the bar weakens or stabilises, the system transitions into the old, quenched state.

The absence of $D_{n}(4000)$ bimodality in barred S0s fits naturally within this framework. These galaxies are already gas-poor and dynamically evolved; their bars persist but no longer influence the star-formation cycle. Consequently, their pseudo-bulges are uniformly old, representing the endpoint of secular evolution. It is plausible that some barred spirals with old pseudo-bulges evolve into barred S0s once disc quenching becomes global \citep[e.g.][]{Vaghmare13, vera16}. In summary, the bimodal $D_{n}(4000)$ distribution—corroborated by the complementary $H\delta_{A}$ behaviour demonstrates that bar-driven secular evolution can both sustain and subsequently quench central star formation, depending on the evolutionary stage of the host galaxy. The implications of this transition are further examined in Section~\ref{sec:color_size} through a comparison of colours and structural parameters of the bulge, bar, and disc components.

\subsection{Colours and effective radii of different galactic components}
\label{sec:color_size}
To explore whether the stellar-age bimodality identified in Section~\ref{sec:agebimodality} is accompanied by systematic variations in the global photometric properties of host galaxies, we compare the colours and structural parameters of the bulge, bar, and disc components derived from the multi-band decompositions of \citet{Kruk18}. These quantities offer insight into how bar-driven secular processes impact both the central and extended regions of galaxies throughout their evolutionary progression.

The left panel of Figure~\ref{fig:col_size} illustrates the $(g-i)$ colour distributions of the bulge, bar, and disc for YB-Sps in blue, OB-Sps in red and OB-S0s in black. For spirals, the colour distributions show a wide spread, with a clear separation between the YB-Sps and OB-Sps defined by $D_{n}(4000)=1.5$. Galaxies hosting young pseudo-bulges are, on average, $\sim0.23$ mag bluer in their bulge colours and $\sim0.14$ mag bluer in the bar compared to those with old pseudo-bulges. Their discs are also systematically bluer, consistent with active star formation throughout the galaxy. By contrast, barred S0s and OB-Sps both display uniformly red bulges, while the discs of S0s and OB-Sps occupy intermediate colours indicative of quiescent or fading stellar populations. Interestingly, the bars in S0s are bluer than OB-Sp and YB-Sp bars, which hints towards the bar rejuvenation in S0s. Recently, \citep{Barway20} has reported the occurrence of the bluer bars in S0 galaxies of \citet{Kruk18} sample and suggested that the environment might have played an important role, since most of these galaxies reside in the intermediate-density environment. In both OB-Sp and OB-S0 subsamples, bulges exhibit the redder colours, followed by bars, with discs being the bluest. This ordering ($g-i$)$_{ bulge}$ $>$ ($g-i$)$_{bar}$ $>$ ($g-i$)$_{disc}$ holds consistently and reflects the well-established inside-out quenching paradigm, where older stellar populations dominate central regions while younger stars and ongoing star formation are preferentially located in discs \citep{Delgado15, Belfiore18}. These colour trends reinforce the age bimodality inferred from $D_{n}(4000)$ and $H\delta_{A}$. In YB-Sps, the bar and bulge share comparable colours, suggesting ongoing or recently terminated star formation within the central kiloparsec. For OB-Sps, both components are considerably redder than the disc, implying that central quenching precedes disc quenching, a characteristic of inside-out or bar-driven suppression of star formation \citep{Lin19, Guo19}.    

A related diagnostic of structural evolution is the relative size of galactic components. The right panel of Figure~\ref{fig:col_size} compares the distributions of effective radii for the bulge ($r_{e,b}$), bar ($r_{e,bar}$), and disc ($r_{e,d}$) in both morphologies. The medians $r_{e,b}$ are comparable between barred spirals and S0s, suggesting that secular processes build pseudo-bulges of similar physical extent regardless of host morphology. However, the bars in S0s are marginally shorter, and their discs are more compact than those in spirals. This difference likely reflects disc fading and contraction following the exhaustion of cold gas. The median value of $r_{e,bar}/r_{e,d}$ is typically smaller in S0s ($\approx0.73$) than in spirals ($\approx0.82$), consistent with the picture that bar-driven inflows redistribute disc material inward, building up the bulge and weakening the outer disc as the system evolves.

In this framework, galaxies hosting young pseudo-bulges represent an early phase in bar-driven evolution: the bar efficiently channels gas inward, sustaining star formation in the central region, with a blue, extended disc. As the gas reservoir is depleted, both the bar and bulge redden, and the disc becomes more compact and less active. This gradual transformation ultimately leads to systems resembling the barred S0s in our sample, which are gas-poor, red, and structurally more concentrated. The continuity of colour and size trends across these populations provides compelling evidence that bar-driven secular evolution not only shapes bulge ages but also governs the broader morphological transition from star-forming spirals to quenched S0 galaxies.

\subsection{Dependence on Stellar Mass}
\label{sec:st_mass}

The role of stellar mass on various properties of galaxies such as colour, size, and star formation rate is well known \citep{Conselice06}. The \citet{Kruk18} sample contains galaxies of stellar masses from 10$^8$ to 10$^{11.5}$ M$_\odot$ and has found that the discs and bars are bluer for low-mass galaxies compared to massive galaxies when using 10$^{10.25}$ M$_\odot$ as mass division. We utilised a similar mass division for our sample galaxies across different populations and plotted the $(g-r)$ colour for disc (left panel), bar (middle panel), and bulge (right panel) in Figure~\ref{fig:mass_cut_col}. In general, we found that high-mass galaxies ($M_* > 10^{10.25} \, M_\odot$) exhibit systematically redder colours in all three components compared to their low-mass counterparts. The shift is particularly pronounced in bars ($\Delta$($g-i$) $\approx$ 0.1–0.3 mag) and bulges ($\Delta$($g-i$) $\approx$ 0.1–0.2 mag), consistent with mass-driven quenching and the buildup of older stellar populations in more massive systems \citep{kauffmann03, Gallazzi05}. This also mirrors the global colour–mass relation in SDSS galaxies \citep{Baldry04} and reflects the transition from star-forming to quenched systems at the characteristic mass scale. 

Unlike the spiral populations, barred S0 galaxies exhibit almost no mass-dependent colour evolution in any structural component. Their discs maintain a constant red colour (($g-i$) = 1.07) from low- to high-mass systems, while bars and bulges become only marginally redder ($\Delta$($g-i$) $\approx$ +0.02 and +0.04 mag, respectively). This stability demonstrates that the stellar populations of lenticular galaxies are already dominated by old stars. We found that in massive galaxies, the stellar bars of barred S0 galaxies are markedly bluer (($g-i$)= 1.11) than those of both YB-Sps (1.24) and OB-Sps (1.32), while in low-mass systems, the S0 bars have a similar colour as OB-Sps. This striking mass-dependent reversal indicates that bar-driven quenching operates far more efficiently above $\sim$10$^{10.25}$M$\odot$, where stronger and faster bars rapidly deplete gas along their length, producing extremely red bars in massive spirals that still retain disc gas \citep{Khoperskov18, Geron21}. Once the galaxy completes its transformation into an S0, the gas-poor bar ages only slowly and can be rejuvenated by late minor mergers or residual central star formation, resulting in significantly bluer colours than observed in massive OB-Sp bars \citep{Eliche-Moral18, Fraser-McKelvie18}. The fact that massive S0 bulges remain red while their bars are blue can also be due to the secular origin of these structures, with boxy/peanut pseudo-bulges and bars containing a substantial intermediate-age stellar population formed during earlier bar-induced gas inflows.

\begin{figure}
    \centering
    \includegraphics[scale = 0.45]{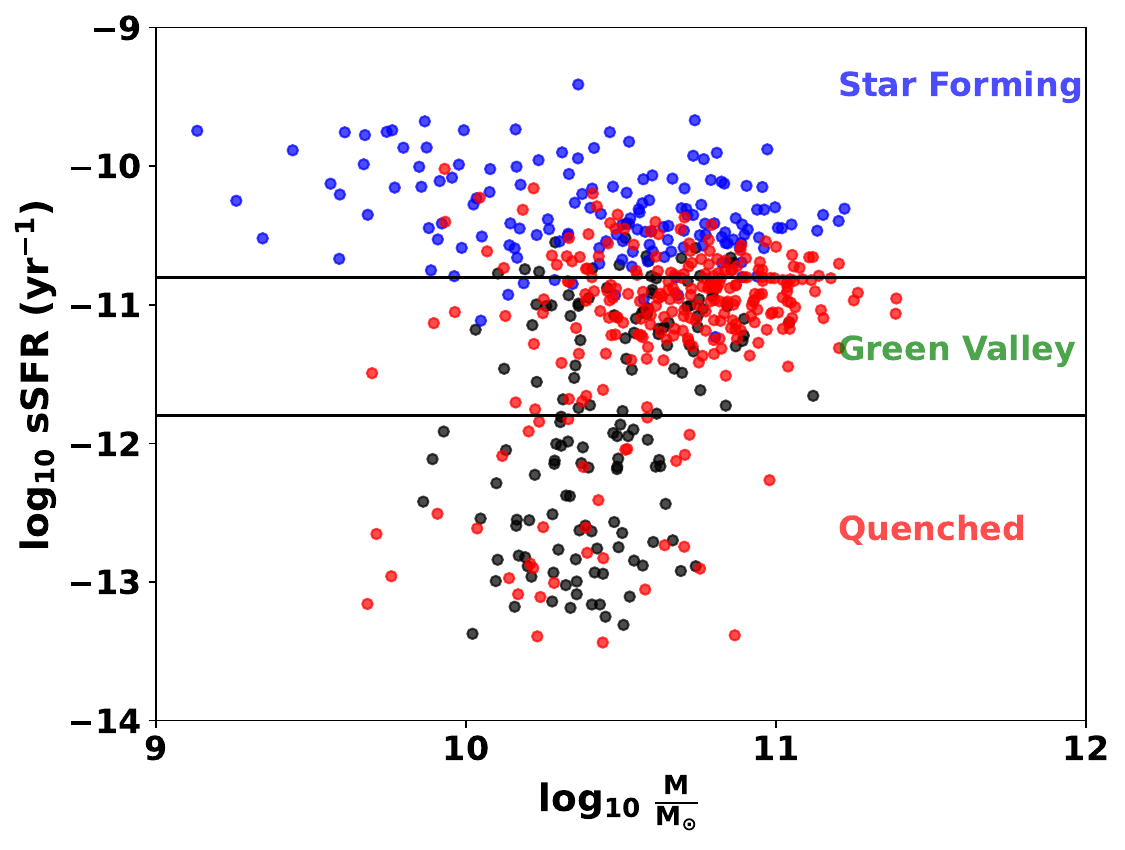}
    \caption{Specific star-formation rate (sSFR) versus stellar mass for barred galaxies hosting pseudo-bulges. Young-bulge spirals are shown in blue, old-bulge spirals in red, and barred S0 galaxies in black. Horizontal dashed lines mark the divisions between star-forming, green-valley, and quenched systems following \protect\cite{Salim14}.}
    \label{fig:ssfr_mass}
\end{figure}

\subsection{Star formation activity}
\label{sec:sf_activity}

Since a galaxy's total gas mass and its global star formation rate are correlated \citep{Kennicutt98}, the stellar population age in pseudo-bulges must be shaped by the star formation rate of a galaxy as a whole. Galaxies with higher SFRs are expected to host younger mean stellar ages in their pseudo-bulges, as elevated gas transport sustains prolonged or recent central star formation activity. To further understand the possible connection and the physical origin of the stellar age bimodality of pseudo-bulges discussed in Section~\ref{sec:agebimodality}, we investigate the total star-formation activity of our sample galaxies using the specific star-formation rate (sSFR) and the stellar mass ($M_{*}$). Figure~\ref{fig:ssfr_mass} shows the distribution of pseudo-bulge hosting galaxies in the sSFR--stellar-mass ($M_{*}$) plane. Both sSFR and $M_{*}$ are taken from GSWLC - X2 \citep{salim18}, having 90\% coverage of SDSS. Out of 677 galaxies, we are able to retrieve mass and SFR for 609 galaxies, detailed classification is given in Table~\ref{tab:ssfr_mass}.

Galaxies are divided into the three categories of star-forming (SF; $\log{\rm sSFR}\!>=\!-10.8$), green-valley (GV; $-11.8<=\log{\rm sSFR}\!<\!-10.8$), and quenched (Q; $\log{\rm sSFR}\!<\!-11.8$) systems, following \citet{Salim14}. YB-Sps ($D_{n}(4000)\!<\!1.5$) occupy predominantly the SF region, whereas both OB-Sps and OB-S0s mostly populate the GV and Q sequences. Nearly 45\% of barred S0s lie firmly within the quenched regime, consistent with their uniformly high $D_{n}(4000)$ values. At fixed mass, OB-Sps exhibit lower sSFRs compared to their young counterparts, and OB-S0s' sSFR is even lower. This scenario emerges as a connected history of star formation activity in the galaxy disc and in the pseudo-bulges across barred spiral and barred S0 galaxies. Thus, the origin of the old stellar population within OB-Sps and OB-S0 galaxies is hypothesized to be a direct result of bar-driven gas dynamics, leading to the eventual shutdown of star formation across the galaxy. The $H\delta_{A}$ index, discussed earlier in Section~\ref{sec:agebimodality}, provides a short-timescale probe of the same phenomenon. When examined as a function of sSFR, $H\delta_{A}$ shows a clear anti-correlation: galaxies with high sSFR exhibit strong $H\delta_{A}$ absorption, while quenched systems show weak Balmer features. This trend demonstrates continuity between the spectroscopic and photometric indicators of recent star formation. Galaxies with young pseudo-bulges, high $H\delta_{A}$, and low $D_{n}(4000)$ occupy the star-forming main sequence, whereas old pseudo-bulge systems with low $H\delta_{A}$ fall below it, reflecting a recent cessation of star formation. Thus, $H\delta_{A}$ and sSFR jointly trace a smooth evolutionary pathway from star-forming to quenched barred galaxies. It is also important to consider the effect of the bar on the age bimodality of pseudo-bulges in these systems. Since pseudo-bulges of unbarred spirals do not show any age bimodality \citep{mishra17a}, the bars must be playing a driving role in the age bimodality of pseudo-bulges in barred spirals.

\begin{figure}
    \centering
    \includegraphics[scale = 0.5]{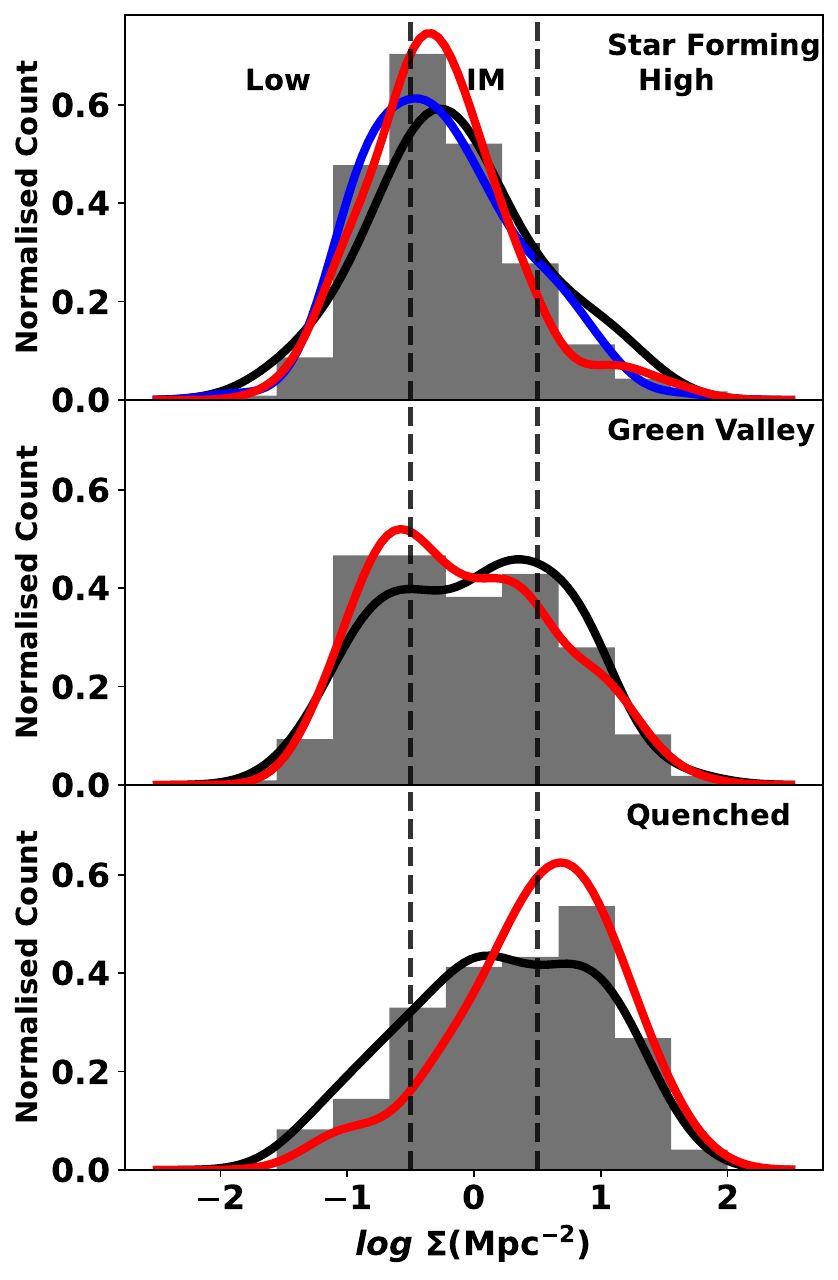}
    \caption{Local environmental density, $\log \Sigma\,(\mathrm{Mpc}^{-2})$, for star-forming, green-valley, and quenched barred galaxies hosting pseudo-bulges. The classification into star-forming, green-valley, and quenched systems is based on the sSFR--$M_{*}$ separation of \citet{Salim14}. Gaussian kernel–smoothed distributions for OB--Sp, YB--Sp, and OB--S0 hosts are shown in red, blue, and black, respectively. The vertical dashed lines divide the sample into low-density ($\log \Sigma < -0.5$), intermediate-density ($-0.5 < \log \Sigma < 0.5$), and high-density ($\log \Sigma > 0.5$) environments.}
    \label{fig:envt-hist}
\end{figure}

Since star formation in galaxies can be affected by both morphology and environment, it is important to investigate the connection between the age distribution of pseudo-bulges and the environment density.
Figure~\ref{fig:envt-hist} shows the distribution of local environmental density ($\Sigma$) for the galaxies in our sample. The top, middle, and bottom panels present the environments of SF, GV, and Q galaxies, respectively, fro YB-Sps, OB-Sps and OB-S0s. Star-forming galaxies for all three subsamples mostly reside in the low- to intermediate-density region. Since most of the YB-Sps are star-forming, we didn't include their distribution for the GV subclass. Galaxies falling in the GV region show a broad range of environments for both OB-Sps and OB-S0s. But the fraction of quenched spirals increases with density. In contrast, quenched OB-S0s reside in the IM to high density region, typical of group or cluster environments. This implies that while bar-driven processes dominate the initial quenching, environmental effects likely accelerate the final transition of YB-Sps to OB-Sps. Thus, environmental effects appear to play a secondary but non-negligible role in the evolution of these systems.

\subsection{Impact of presence of AGN}
\label{sec:agn}
The evolution of galaxies is also regulated by the growth of their central supermassive black holes (SMBHs). Gas inflow toward the nucleus can trigger accretion onto the SMBH and ignite an active galactic nucleus (AGN) \citep{Ellison11}. During such phases, AGN can return substantial energy to the surrounding interstellar medium, heating or expelling gas and thereby suppressing further inflow and central star formation, a process widely referred to as AGN feedback. By preventing gas from cooling and accumulating in the inner kiloparsec, AGN feedback can effectively quench star formation both in the central region and along the bar \citep{Bluck20, Lammers23}. Observational studies indicate that the AGN fraction increases toward the transition regime between star-forming and quiescent galaxies \citep{Schawinski07, Schawinski09}, suggesting a role for AGN in mediating this evolution. Complementary work further indicates that large-scale bars may facilitate SMBH growth by funnelling gas toward the nucleus, and that barred galaxies are more likely to host AGN than their unbarred counterparts \citep{garland24, marels25, lamarca26}.

\begin{figure*}
    \centering
    \includegraphics[scale = 0.38]{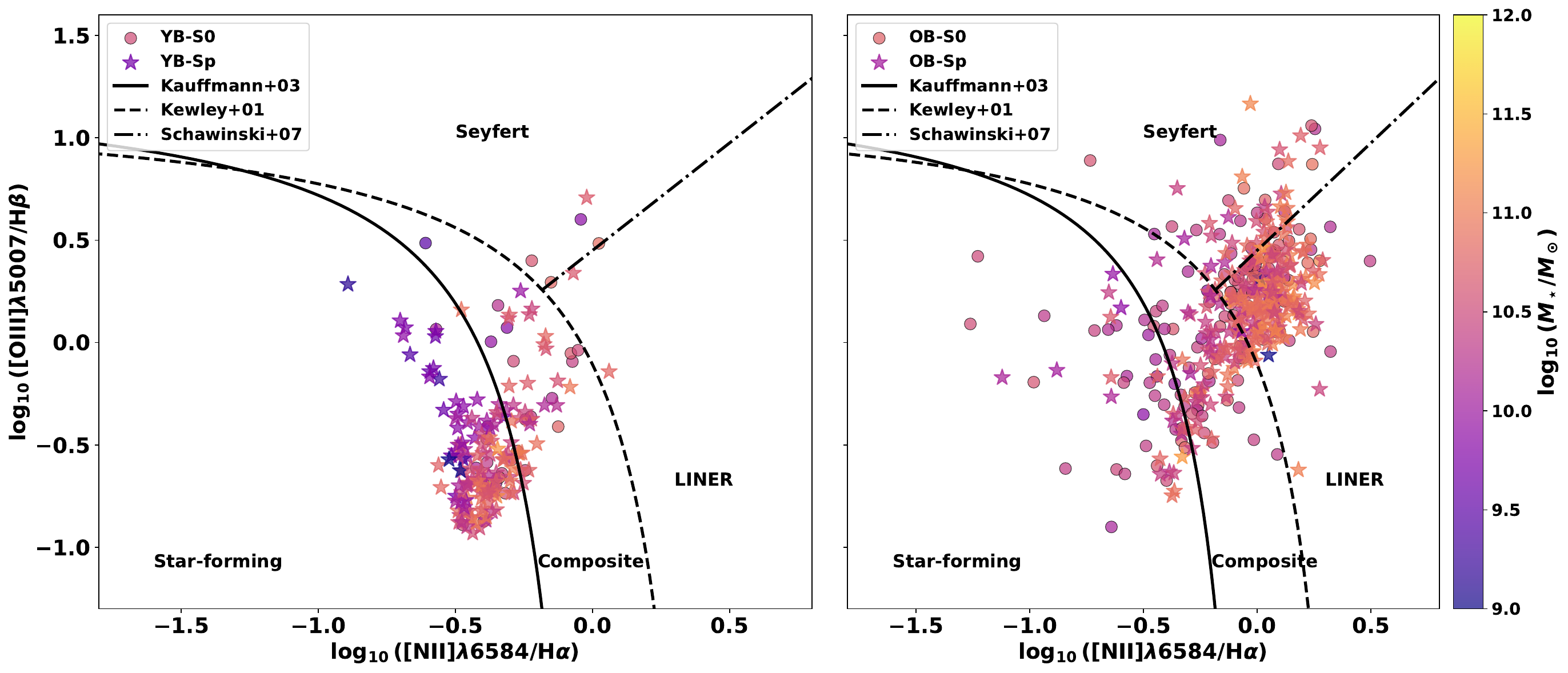}
    \caption{BPT diagnostic diagrams showing $\log\big([\mathrm{O\,III}]\,\lambda5007/\mathrm{H}\beta\big)$ versus $\log\big([\mathrm{N\,II}]\,\lambda6584/\mathrm{H}\alpha\big)$ for the YB (left) and OB (right) subsamples. Individual galaxies are colour-coded by stellar mass. Classification curves from \citet{Kauffmann03b}, \citet{Kewley01}, and \citet{Schawinski07} are overlaid to separate star-forming, composite/LINER, and Seyfert-like ionisation regimes.}
    \label{fig:agn}
\end{figure*}

To identify the dominant ionisation source in our sample, we used the Baldwin–Phillips–Terlevich (BPT) diagnostic \citep{Baldwin81}, which classifies emission mechanisms based on characteristic nebular line ratios. We adopt the standard [O\,\textsc{iii}]~$\lambda5007$/H$\beta$ versus [N\,\textsc{ii}]~$\lambda6584$/H$\alpha$ flux ratios from the MPA–JHU measurements of the SDSS spectra. Figure~\ref{fig:agn} presents the BPT distributions for our pseudo-bulge subsamples, with YB hosts shown on the left and OB hosts on the right. The solid division in each panel \citep{Kauffmann03b} marks the locus of star-forming galaxies. In contrast, the dashed curve \citep{Kewley01} defines the theoretical upper envelope for pure star formation, above which AGN or other non-stellar processes must drive ionisation. The dotted–dashed line \citep{Schawinski07} separates Seyfert and LINER-like AGN. The number of galaxies in each region is listed in Table~\ref{tab:ssfr_mass}.

From this classification, it is evident that the majority of YB hosts lie below the \citet{Kauffmann03b} line, consistent with ionisation dominated by star formation, with only a small fraction hosting Seyfert\,2 activity. In contrast, most OB–Sp and OB–S0 systems fall above the \citet{Kewley01} boundary, indicating excitation mechanisms other than star formation. These are predominantly LINER-like, with a minority showing Seyfert\,2 signatures. This agrees with the trends reported by \citet{Masters11}, who found that red spirals more frequently host LINER+Seyfert emission than blue spirals, with incidence increasing with stellar mass. Because LINER-like emission can also arise from photoionisation by evolved stellar populations \citep{Stasi08, Cid11}, its dominance in OB hosts likely reflects their older central stellar populations rather than widespread AGN activity.


\begin{table}
\centering
\begin{tabular}{lrrr@{}rrrrr}
\toprule
      & \multicolumn{3}{c}{\textbf{Morph}} 
        & \multicolumn{5}{c}{\textbf{BPT Classifaction}} \\

\cmidrule(lr){2-4} \cmidrule(lr){6-9}
      & \textbf{SF}  & \textbf{GV} & \textbf{Q}   & & \textbf{SF}  & \textbf{Composite} & \textbf{LINER} & \textbf{Seyfert} \\ 
\midrule
YB-S0 & 14.6  & 2.7   & 1.1   & & 8.3   & 8.3        & 00     & 2.4 \\
OB-S0 & 9.2  & 33.5  & 39  & & 18.3  & 21.3       & 29    & 12.4 \\
YB-Sp & 33.7 & 2.6  & 00  & & 29.6  & 7.4  & 0.5 & 0.25 \\
OB-Sp & 17  & 38.4 & 8.3  & & 5.7   & 14.5 & 33.7  & 8.4 \\
\bottomrule
\end{tabular}
\caption{Fraction of galaxies in each subsample (spirals and S0s), classified according to their location in the sSFR--$M_{*}$ plane as star-forming (SF), green-valley (GV), or quenched (Q). 
This classification is available for 609 systems in total (424 spirals and 185 S0s). For the same subsamples, we also list the fractions based on BPT emission-line diagnostics, separating galaxies into star-forming, composite, LINER, and Seyfert classes. 
Reliable emission-line measurements are available for 575 galaxies (406 spirals and 169 S0s). Pseudo-bulge hosts are further divided into young (YB; $D_{n}(4000) \leq 1.5$) and old (OB; $D_{n}(4000) > 1.5$) populations.}
\label{tab:ssfr_mass}
\end{table}

\section{Discussion \& Conclusion}
\label{sec:summary}
In this study, we have investigated the stellar populations, photometric properties, and star-formation activity of a sample of barred galaxies drawn from the SDSS DR7, with a focus on systems hosting pseudo-bulges. By dividing the sample into barred spirals and barred S0s, we examined how bar-driven secular processes influence the evolution of central stellar populations and galaxy morphology. The results presented in Sections~\ref{sec:agebimodality}--\ref{sec:agn} reveal that pseudo-bulges in barred galaxies exhibit a pronounced stellar-age bimodality, with peaks at $D_{n}(4000)\!\approx\!1.3$ and $\approx\!1.8$, whereas barred S0s show higher $D_{n}(4000)$, dominated by old stellar populations. This is closely linked to the presence and evolutionary state of the bar and can be understood within the framework of bar-driven secular evolution, wherein the bar both promotes and suppresses central star formation depending on the gas content and dynamical evolution of the host galaxy. 

It is instructive to place these results in the broader context of pseudo-bulge evolution across different galaxy populations. Previous studies have shown that pseudo-bulges in unbarred spiral galaxies exhibit largely unimodal age distributions, whereas S0 galaxies display clear evidence of bimodality \citep{mishra17a}. Our results demonstrate that pseudo-bulges in barred spiral galaxies also exhibit a pronounced bimodality in their stellar populations. This indicates that bimodality is not unique to barred systems, but can arise through multiple evolutionary pathways.

During the early, gas-rich phase of bar evolution, gas inflow triggers circumnuclear starbursts and gradually builds a pseudo-bulge through dissipative processes \citep{kormendy04}. The YB-Sps in our sample, characterised by low $D_{n}(4000)$ values, strong $H\delta_{A}$ absorption, blue bulge and bar colours, and elevated specific star-formation rates, fit this picture. Their properties imply active or recently enhanced central star formation, likely sustained by continuing bar-driven inflows. The distribution of these systems along the star-forming main sequence suggests that the bar operates as a secular regulator of gas inflow rather than a transient instability. These galaxies are also mostly found in low-density environments, intimating the bar-driven secular evolution of YB-Sp galaxies.

In contrast, the OB-Sps with high $D_{n}(4000)$ and weak $H\delta_{A}$ absorption show clear evidence of quenched central regions. Numerical simulations predict that as the bar strengthens and the central mass concentration increases, resonances and vertical heating suppress further inflow, which stabilises or exhausts the gas reservoir \citep[e.g.][]{Fanali2015}. We also observed a correlation between colour and structure of the morphological components, which is in line with previous studies of barred galaxies \citep[e.g.][]{gadotti09, Masters11, vera16}. \citet{Masters11} found that the fraction of strong bars increases among redder, more massive discs, while \citet{vera16} showed that barred galaxies with redder bulges tend to have smaller, less star-forming discs. Our findings extend this picture by demonstrating that the colours of the bulge and bar correlate directly with the stellar ages inferred from spectroscopic indices. The close correspondence between bulge and bar colours suggests that both components evolve synchronously, likely through the same secular processes. A recent simulation study by \cite{frosst25} shows that bars can induce central quenching by accelerating the growth of the supermassive blackhole, which leads to feedback and eventually cessation of star formation. From the BPT classification of our sample galaxies, we found that only a small fraction of OB-Sps and OB-S0s host active AGN (Seyfert type 2), which can influence the cessation of star formation in these galaxies. A recent study of triple AGN host barred interacting galaxy group by \citet{Keshri25} have found that ionization from an active AGN in NGC 7733 influences the star formation in the central region and leads to the bar quenching. However, the galaxy also hosts a long and strong bar along with a nuclear disc and pseudo-bulge, which further suggests that bar-driven gas inflow triggers the star formation in the central region and activates the AGN. In this study, since very few galaxies host an active AGN, the bimodality seen in the pseudo-bulges is likely due to bar-driven processes.
Thus, the resulting reduction in star formation leads to the formation of a dynamically evolved, red pseudo-bulge. These OB-Sps also host red and long bars. Observational studies have shown that strong bars are more common in redder, more massive systems \citep{Masters11, vera16}, consistent with our findings, supporting the idea that bars not only initiate bulge growth but also accelerate its quenching. The coexistence of YB-Sps and OB-Sps, therefore, represents different stages along a single bar-regulated evolutionary sequence. Most of these OB-Sps are members of the GV population, suggesting an evolutionary transition of SF YB-Sps to GV OB-Sps. OB-Sps are found in IM and high-density regions, where the fraction of quenched galaxies increases with environmental density. This suggests that external gas depletion or environmental quenching complements bar-driven secular evolution. 

Barred S0 galaxies appear to mark the terminal phase of this process. Their uniformly high $D_{n}(4000)$ indices, low $H\delta_{A}$, and red bulge and disc colours indicate fully quenched central regions. Structurally, they are nearly indistinguishable from old pseudo-bulge spirals, supporting the view that gas-poor barred spirals fade into S0s once star formation ceases across the disc \citep{Vaghmare18}. These galaxies mostly reside in group- or cluster-like environments, where ram-pressure stripping or strangulation can further hasten gas removal and disc fading, thereby reinforcing the morphological transformation. We note, however, that the presence of a bimodal age distribution is not unique to barred systems, as similar behaviour has been reported in pseudo-bulges of unbarred S0 galaxies \citep{mishra17a}. This suggests that while bars provide an efficient internal pathway for regulating central star formation in spirals, environmental processes and gas depletion can also drive the evolution of pseudo-bulges in S0 galaxies.

An interesting exception to this quenching pathway is provided by S0 galaxies with unusually blue or star-forming bars, interpreted as evidence of ``bar rejuvenation'' \citep{Barway20}. These systems, typically found in group environments, may have experienced minor interactions or external gas accretion events that briefly reignite star formation along the bar without rebuilding a star-forming disc \citep[e.g.][]{Kaviraj09, Salim12, rathore22}. Such rejuvenation episodes are expected to be more readily detectable in gas-poor S0 galaxies, where even modest amounts of recent star formation can significantly alter observed colours, whereas in spirals, ongoing star formation may mask similar signatures.  These events appear to be relatively rare and short-lived, representing perturbations to, rather than reversals of, the overall quenching pathway. They nevertheless highlight the complex interplay between internal secular processes and external environmental effects in shaping the evolution of barred galaxies.

Recent JWST studies reveal that bars are already present, though rare, at $z\!\sim\!3$--4 and steadily increase in frequency, size, and strength toward the local Universe, implying that bar-driven secular evolution becomes progressively more important with cosmic time \citep{Guo2025, Geron2025}. The rising bar fractions in massive field disc galaxies in the local universe can provide the high-redshift comparison for the stellar-population bimodality observed in local barred spirals, i.e. young, blue pseudo-bulges versus old, quenched pseudo-bulges. A recent study by \citet{Huang25} reveals evidence of bar-driven gas inflows toward the galactic centre, fuelling intense central star formation in a galaxy observed only 2.8 billion years after the Big Bang. These findings indicate that bar-driven secular evolution was already active in the early Universe. The JWST detection of early bars suggests that the younger pseudo-bulges correspond to systems where bar-driven inflows remain active, consistent with the intense central star formation seen in our low-$D_{n}(4000)$ population. Conversely, the persistence of strong bars across cosmic time, together with their growth toward lower redshift, supports a scenario in which bars later suppress star formation by stabilising or depleting central gas, resulting in the old, quenched pseudo-bulges that dominate barred S0 galaxies. Environmental results from JWST cluster studies \citep{MendezAbreu2023} further show that bar formation is inhibited in low-mass cluster discs, explaining why quenched pseudo-bulges in S0 galaxies are primarily associated with massive, long-lived bars. Together, these findings suggest that bars drive a two-phase bulge-evolution pathway, characterised by early growth through gas inflow and later quenching via secular dynamical processes, thereby linking high-$z$ barred progenitors to the diverse pseudo-bulges observed today.

IFU studies of nearby barred galaxies, particularly from the TIMER survey using MUSE \citep{Gadotti2020, deSaFreitas2023}, reveal that bar-driven gas inflow towards the central region results in the formation of rotationally supported nuclear discs, whose sizes, rotation support, and stellar populations scale with bar length and evolutionary stage. These findings align closely with our analysis. TIMER results show that older, longer bars typically host larger and more evolved nuclear discs, supporting a secular evolutionary sequence in which bars first funnel gas inward to assemble a young pseudo-bulge, and subsequently stabilise or deplete the central gas reservoir, leading to quenching. A recent study by \citet{Le_Conte26} reports the discovery of a nuclear disc at Cosmic Noon, providing compelling evidence for bar-driven galaxy evolution at early epochs. The presence of ongoing star formation and comparatively younger stellar populations within the nuclear disc strongly indicates efficient bar-driven gas inflow. These findings reinforce the role of bars in accelerating the secular evolution and fast maturation of galaxy discs and bars at Cosmic Noon. In this framework, the bimodal pseudo-bulge age distribution we observe naturally reflects distinct phases of bar-driven evolution: an early growth phase and a later quenching phase. Combined with recent JWST detections of bars out to $z\!\sim\!4$, which indicate that bars become increasingly common and dynamically influential over cosmic time, a coherent picture emerges in which bar-driven inflow, nuclear disc formation, and eventual central quenching together shape the pseudo-bulge population observed in the local Universe.

Taken together, the evidence indicates that bars play a dual role in galaxy evolution: they first act as conduits for inflowing gas that stimulate central star formation and pseudo-bulge growth, and subsequently become agents of quenching once the inflowing material is consumed or dynamically stabilised. When combined with environmental gas depletion, this process naturally links gas-rich barred spirals hosting young pseudo-bulges to gas-poor barred S0s with old, passively evolving bulges. Bars thus emerge as key drivers of both the formation and cessation of central star formation, providing a coherent secular pathway connecting blue, star-forming discs to red, quenched S0 galaxies. Testing this evolutionary framework requires spatially resolved observations of stellar populations and gas kinematics. Integral-field spectroscopy observations from MaNGA, SAMI, and MUSE can trace stellar age and metallicity gradients across bars and bulges, directly probing inflow-driven star-formation histories. Complementary molecular gas mapping with ALMA will quantify the gas content and dynamics in both star-forming and quenched bars. Together, these datasets will determine whether bar-induced quenching proceeds via gradual gas exhaustion or dynamical stabilisation, thereby constraining the full timeline of secular evolution of pseudo-bulges from blue barred spirals to red barred S0 galaxies.


\section{Acknowledgements}
We thank the anonymous referee whose thoughtful and constructive comments have greatly helped in improving the presentation of this paper. KK acknowledges the support from UWA – ICRAR Higher Degree by Research Scholarship and the Scholarship for International Research Fees from the University of Western Australia. KK thanks the Indian Institute of Astrophysics, Bengaluru, for providing the opportunity and resources to work on this project in April to June 2023. KK also thanks the valuable comments provided by Prof. Aaron Robotham.

Funding for the SDSS and SDSS-II has been provided by the Alfred P. Sloan Foundation, the Participating Institutions, the National Science Foundation, the U.S. Department of Energy, the National Aeronautics and Space Administration, the Japanese Monbukagakusho, the Max Planck Society, and the Higher Education Funding Council for England. The SDSS Web Site is http://www.sdss.org/.
      
The SDSS is managed by the Astrophysical Research Consortium for the Participating Institutions. The Participating Institutions are the American Museum of Natural History, Astrophysical Institute Potsdam, University of Basel, University of Cambridge, Case Western Reserve University, University of Chicago, Drexel University, Fermilab, the Institute for Advanced Study, the Japan Participation Group, Johns Hopkins University, the Joint Institute for Nuclear Astrophysics, the Kavli Institute for Particle Astrophysics and Cosmology, the Korean Scientist Group, the Chinese Academy of Sciences (LAMOST), Los Alamos National Laboratory, the Max-Planck-Institute for Astronomy (MPIA), the Max-Planck-Institute for Astrophysics (MPA), New Mexico State University, Ohio State University, University of Pittsburgh, University of Portsmouth, Princeton University, the United States Naval Observatory, and the University of Washington.

\section*{Data Availability}
 This work uses data from SDSS, MPA-JHU, GSWLC and the catalogue by \cite{Kruk18}, all available publicly online.
 


\bibliographystyle{mnras}
\bibliography{paper} 

\bsp	
\label{lastpage}
\end{document}